\documentclass[aps,twocolumn,floatfix,rspublic]{revtex4-1} 

\usepackage{amsmath,amsfonts,amssymb,graphicx,color,bbm}
\usepackage{hyperref}

\begin{document}

\title{Coherence and Decoherence in Biological Systems: Principles of Noise
Assisted Transport and the Origin of Long-lived Coherences}
\author{A.~W. Chin, S.~F. Huelga and M.~B. Plenio}
\address{Institute of Theoretical Physics, Universit{\"a}t Ulm,
Albert-Einstein-Allee 11, 89069 Ulm, Germany}

\begin{abstract}
The quantum dynamics of transport networks in the presence of noisy
environments have recently received renewed attention with the discovery of
long-lived coherences in different photosynthetic complexes. This experimental
evidence has raised two fundamental questions: Firstly, what are the mechanisms
supporting long-lived coherences and secondly, how can we assess the possible functional
role that the interplay of noise and quantum coherence might play in the
seemingly optimal operation of biological systems under natural conditions?
Here we review recent results, illuminate them at the hand of two paradigmatic
systems, the Fenna-Matthew-Olson (FMO) complex and the light harvesting
complex LHII, and present new progress on both questions. In particular we 
introduce the concept of the phonon antennae and discuss the possible microscopic 
origin or long-lived electronic coherences.
\end{abstract}

\maketitle

\section{Background \& Motivation}
Quantum mechanics provides the natural laws that govern the dynamical evolution of
atoms and molecules. Under well controlled conditions in almost
perfectly isolated systems, quantum coherence and entanglement can exist and be
manipulated. While there is a clear notion of sub-systems when individual units
are spatially well separated, as is usually the case in systems suitable for
quantum information processing, this is not necessarily so in molecular complexes,
where partitioning the total systems by means of the choice of appropriate subsystems
is, to some extent, arbitrary. In such situations, when the system is static,
coherence and entanglement can always be made to vanish in a suitably chosen
basis \cite{PlenioV07,Mukamel2010}. Hence, to observe non-trivial quantum effects we need
to force the system out of its preferred eigenbasis, through intervention from
the outside, and thus probe quantum coherences between eigenstates, as well as
the quantum coherence properties of their consequent dynamics. We can then argue
that, while the {\em static} system can {\em sustain} quantum coherence, it is the
dynamics generated by the external perturbation which results in relevant
quantum phenomena and provides the means to probe for interesting quantum
properties.

The external perturbation may either be coherent and controlled, for example
through the action of laser fields, or it may be caused by an interaction
with an unobserved environment. The latter will never be completely avoidable,
as perfect isolation of a physical system from its environment is impossible.
In fact, in bio-molecular complexes the interaction with uncontrolled environments
such as vibrations of the surrounding protein matrix, for example in the form of thermal
fluctuations, is an important driving force of its energetic dynamics and hence its
functionality. One may therefore expect that the interplay between the
internal coherent quantum dynamics of the system and the unavoidable presence
of noise introduced by the environment has been optimized by Nature during evolution and may
be significant for its function.

Exploring to what extent coherent quantum dynamics play a role in composite
quantum systems in contact with environments, such as bio-molecular complexes
and, importantly, what role, if any, genuine quantum traits may play  is thus
an interesting and timely problem. Exploring these issues becomes particularly
challenging in those bio-molecular systems where the strength and timescales
associated with the environment-induced dynamics dynamics are often comparable
to the coherent (Hamiltonian) intra-system dynamics, and cannot therefore be
treated separately.

The investigation of these types of questions has gained further urgency with
the recent discovery of experimental evidence that excitation energy transport
dynamics in the Fenna-Matthew-Olson (FMO) complex and other photosynthetic
aggregates exhibits surprisingly long-lived coherence  \cite{EngelCR+2007,MercerEK+2009,PanitchayangkoonHF+2010}. The FMO complex is an
example of a pigment-protein complex, a network through which electronic excitations
on individual pigments can migrate via excitonic couplings. These experiments suggest
the existence of significant quantum coherences between multiple pigments through the
presence of oscillations in the cross-peaks of 2-D spectra which have been observed
to persist on timescales up to around a picosecond \cite{PanitchayangkoonHF+2010}.
This is notable given that the entire excitation energy transport in the FMO complex
is concluded in less than $5$ps.

Theoretical investigations of the role of pure dephasing noise in
excitation-energy transfer have found that this noise mechanism has
the ability to enhance both the rate and yield of excitation-energy
transfer when compared to perfectly coherent evolution
\cite{MohseniRL+2008,PlenioH2008}. An often quoted argument states that
in the exciton basis typically used in previous studies \cite{ChoVB+2005},
noise-assisted transport is trivially explained as the result
of noise-induced transitions between exciton eigenstates, which cause
an energetic down-hill relaxation towards the reaction centre. While this
argument can serve to suggest that noise-assisted processes might exist in
excitation-energy transfer and other forms of transport in quantum networks,
it falls short, as it cannot explain why
in some networks noise is exclusively detrimental while in others it can
support transport. Answering such a question requires a more detailed study
that identifies basic dynamical principles of quantum networks in contact
with environments.

Here we would like to provide a fresh look at the problem by elucidating
the basic mechanisms which give rise to noise-assisted transport and identifying
where noise and coherence respectively play a constructive role
\cite{MohseniRL+2008,PlenioH2008,OlayaCastroLO+2008,CarusoCD+2009,ChinDC+2010,RebentrostMK+2009,CarusoCD+10}.
Identifying and understanding these mechanisms at a deep, intuitive level
does provide additional value even if the individual processes might have
been known in some previous contexts. Such classifications pave the way
towards tailored experimental tests of the contributions of individual processes
\cite{CarusoMC+2011}. Furthermore, quantum coherence can only be expected
to be relevant if it is sufficiently long-lived. Present experiments suggest
lifetimes of coherences between excited states extending up to the picosecond
range while at the same time the coherence between ground and excited states
are known to last for only about 50-150 fs or so. The origin of these long-lived
coherences remains to be fully clarified. A complete understanding of these fundamental
dynamical principles and the origin of the long-lived coherences will allow
for the optimization of network architectures in order to utilize quantum
coherence and noise to enhance the performance of artificial nano-structure
devices.
\section{Principles of noise assisted transport}
In this section we will briefly review the basic dynamical principles for
noise assisted transport following and extending the ideas developed in
\cite{PlenioH2008,CarusoCD+2009,ChinDC+2010,CarusoCD+10,HuelgaP11}, and further expand
them with the new concept of the \emph{phonon antenna} by which coherent
dynamics may assist the fine-tuning of the network to the spectral density
of the fluctuating environment.

\subsection{Basic dynamical principles}
We have argued that both noise {\em and} coherence are important for the time evolution
of quantum networks in the presence of an environment. But, how does the interplay
between these processes occur?

\subsubsection{Bridging energy gaps \& blocking paths}
Pigment-protein complexes will consist of a number of sites whose energies
will generally exhibit a certain degree of static disorder, i.e. their
on-site energies will differ from site to site. If this energy difference
is larger than the intersite hopping matrix element in the relevant Hamiltonian,
then transitions will be strongly suppressed. Dephasing noise can come to
the rescue here as it will lead to line broadening thus leading to increased
overlap between sites without the loss of excitations from the system (see lhs
of Fig. \ref{Fig1}). Alternatively, one may view dephasing noise as arising
from the random fluctuations of energy levels. As a consequence, the fluctuating
energy levels will occasionally come energetically close enough to allow fast
excitation energy transfer between the sites (see rhs of Fig. \ref{Fig1}).
A moderate amount of fluctuations serves to enhance the transport while excessively strong
fluctuations of the site energies will make the probability smaller for the
sites to be energetically close. Hence we expect an optimal finite noise strength
that maximizes transport between two sites.
\begin{figure}[hbt]
\centerline{\includegraphics[width=8cm]{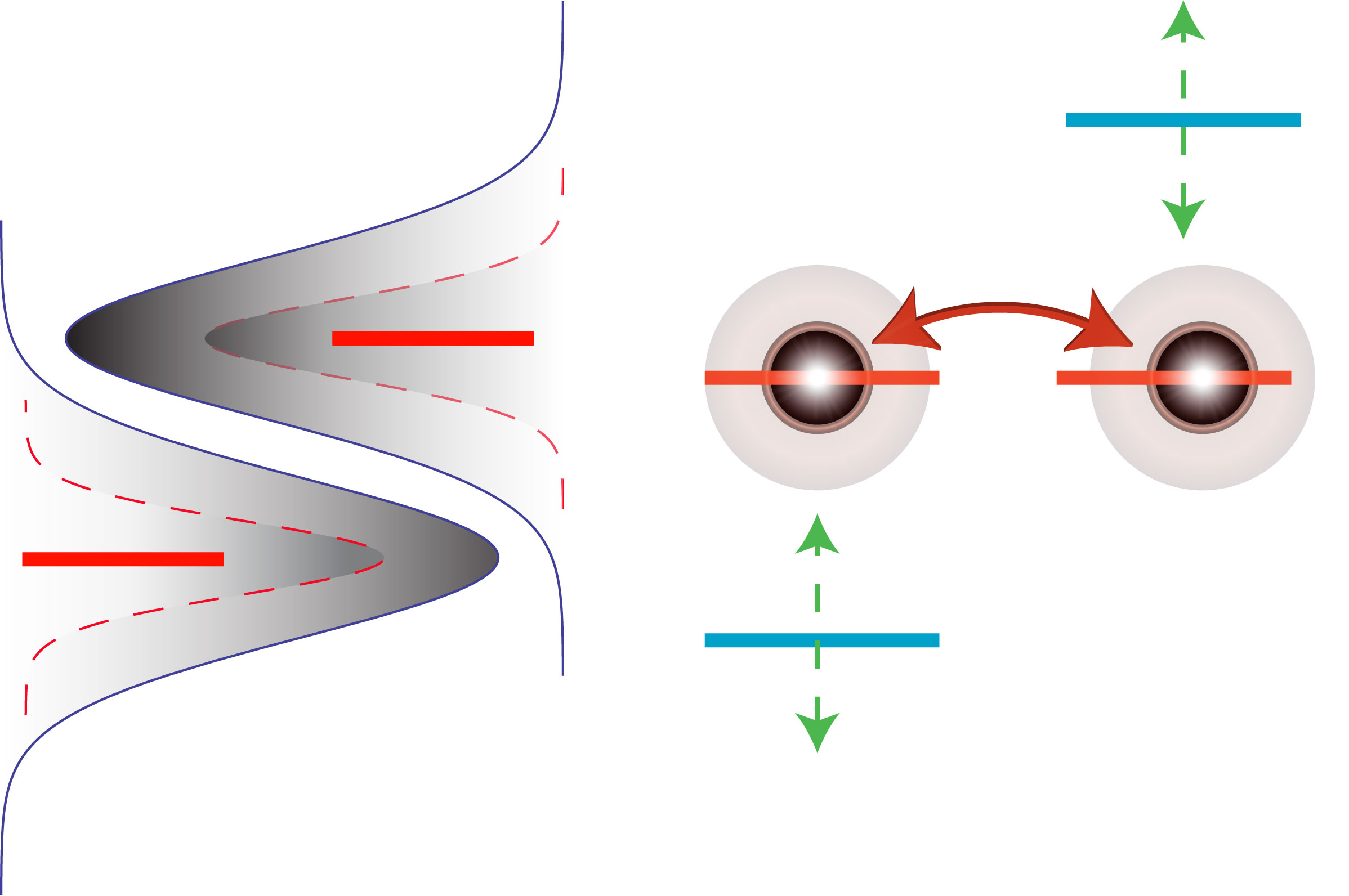}}
\caption{
Left: Local dephasing, for example due to random fluctuations of the energy
levels, leads to line-broadening and hence increased overlap between
sites. Right: Viewing these fluctuations dynamically, one finds that
the energy gap between levels varies in time. The resulting nonlinear dependence
of the transfer rate on the energy gap may therefore lead to an
enhancement of the average transfer rate in the presence of dephasing
noise.
}
\label{Fig1}
\end{figure}
However, the application of excessive noise and the concomitant reduction of the
effective transition rate between sites can also be of advantage as it
may effectively block unfavorable transfer paths from being
followed \cite{ChinDC+2010}.

\subsubsection{Constructive \& destructive interference}
A quantum dynamical system may exhibit a wealth of constructive and destructive
interference effects. The essential nature of this type of effect may be seen in
the simple network depicted in Fig.\ref{Fig1}, where the coherent interaction is
described by a Hamiltonian, $H = \sum_{k=1}^{3} E_i|i\rangle\langle i| +
\sum_{k=1}^2 J_{k3} (|k\rangle\langle 3| + h.c),$ where $|i\rangle$
corresponds to an excitation in site $i$ and we assume $J_{13}=J_{23}$. An excitation
initially prepared in the antisymmetric state $|\psi\rangle = (|1\rangle - |2\rangle)/\sqrt{2}$
forms an eigenstate of this Hamiltonian which has zero overlap with the site 3
which we assume to be coupled dissipatively to a reaction center. Under natural
conditions a pigment-protein complex is not normally excited in such an antisymmetric
state, but will tend to receive a single excitation locally, for example on site $1$.

The initial state localised on site $1$ may of course be considered to be an equally
weighted coherent superposition of the symmetric and the anti-symmetric states, i.e.
$|1\rangle = [(|1\rangle - |2\rangle)/\sqrt{2} + (|1\rangle + |2\rangle)/\sqrt{2}]/\sqrt{2}$.
Thanks to constructive interference the symmetric part of the initial state will
propagate very rapidly into site $3$, and from there into the reaction center,
while the antisymmetric part will not evolve at all. Hence the transfer efficiency
is limited to $50\%$ in this situation. This trapping of population will be suppressed
either via energetic disorder, which will release trapped population by inducing
coherent oscillations between symmetric and anti-symmetric states, or by environmental
dephasing noise which degrades interference effects and therefore destroys coherent
trapping.

We note that these two mechanisms of noise-assisted transport are essentially independent
of microscopic noises models, and could even be seen as highlighting the processes which
noise sources \emph{must} effect in order to drive efficient transport in networks where
disorder and/or interference are important. The precise parameters and dynamical evolutions
by which this is achieved by an environment \emph{do} of course depend on the noise models
and simulation techniques employed, but the physical mechanisms outlined above, which underlie
the advantages and optimality of noise-assisted transport, are \emph{not} qualitatively altered
by any choice of model or approach. As a result, it is reasonable to illustrate the action of
these mechanisms using fairly simple simulation tools, and in section \ref{fmo} we will demonstrate
noise assisted transport using Lindblad master equations. This approach neglects several important
factors, such as the effects of finite temperatures, but for the non-equilibrium conditions we
shall consider, this approach yields results which both demonstrate the dramatic action of
noise-assisted transport and which are also robust - at the same qualitative level of microscopic
description as the proposed mechanisms - to changes in noise model.

Realistic systems require much more description, and simulations of experimental results
certainly require more sophisticated models and simulation techniques. Our aim in this paper
is to slowly build up a picture of the rich array of noise-driven processes which can lead to
efficient energy transport, starting from the most basic ones described above and then incrementally
adding additional details which lead to new mechanisms. As layers are added, so the methods employed
will also become more involved. A good example of this is given in the next section, where we present
a new principle of noise assisted transport which requires a consideration of two important properties
of system-environment interactions which we have so far neglected in this article; the spectral density
of environmental fluctuations and coherent vibronic couplings.


%
\begin{figure}[hbt]
\centerline{\includegraphics[width=7cm]{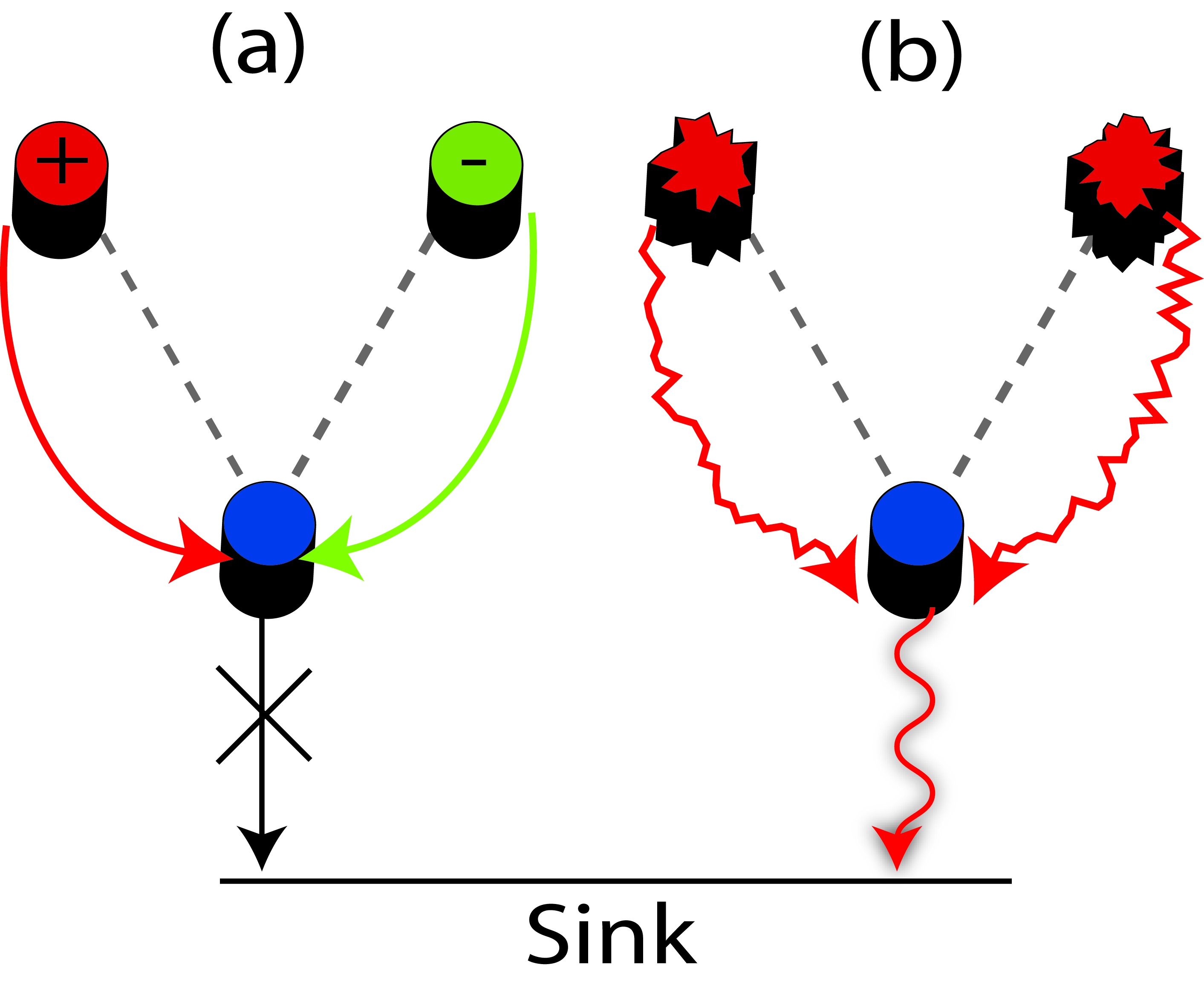}}
\caption{
A three-site network in which two sites $1$ and $2$ are each coupled to a third site $3$ via an
exchange interaction of the same strength. Site three is irreversibly connected to a sink. In (a)
the excitation is delocalized over two sites (red and green) with equal probability of being found
at either site but with a wave function that is antisymmetric with respect to the interchange of red
and green. This state will not evolve due to destructive interference and hence no excitation will
ever reach the reaction center. In (b) pure dephasing causes the loss of phase coherence and the two
tunneling amplitudes no longer cancel, eventually leading to a complete excitation transfer to the sink.}
\label{Fig1}
\end{figure}

\subsubsection{Splitting energy levels - The phonon antenna}\label{phonon}
The presence of strong quantum coherent dynamics between energy levels can
also allow an open system to optimize the efficiency of the energy transport.
To this end, it should be noted that the energy levels of two sites that are
coupled coherently will split, leading to new eigenstates of the global system, one of
which is shifted upwards and one that is shifted downwards. As a consequence,
one of the dressed energy levels may be moved closer to the desired final point of
transport such as the reaction center. As the other level is moved upwards,
a further process is required to drive its population towards the lower lying energy levels.

To this end, note that in this new basis dephasing noise will now induce
transitions between these eigenstates (phase noise in the site basis becomes amplitude noise in
the dressed basis) leading to energy transport towards
the lower lying of the two energy levels. The transition rate between these
two states will depend on
temperature, the matrix element between these two eigenstates {\em and}, crucially, the
spectral density at the energy difference between the two eigenstates.
Matching the energy level splitting to the maximum of the spectral
density of the environmental fluctuations can thus optimize energy
transport. In this sense, we can argue that the two eigenstate of the coupled Hamiltonian
harvest environmental noise to enhance excitation energy transport through
the formation of a "phonon antenna".

\begin{figure}[hbt]
\centerline{\includegraphics[width=8cm]{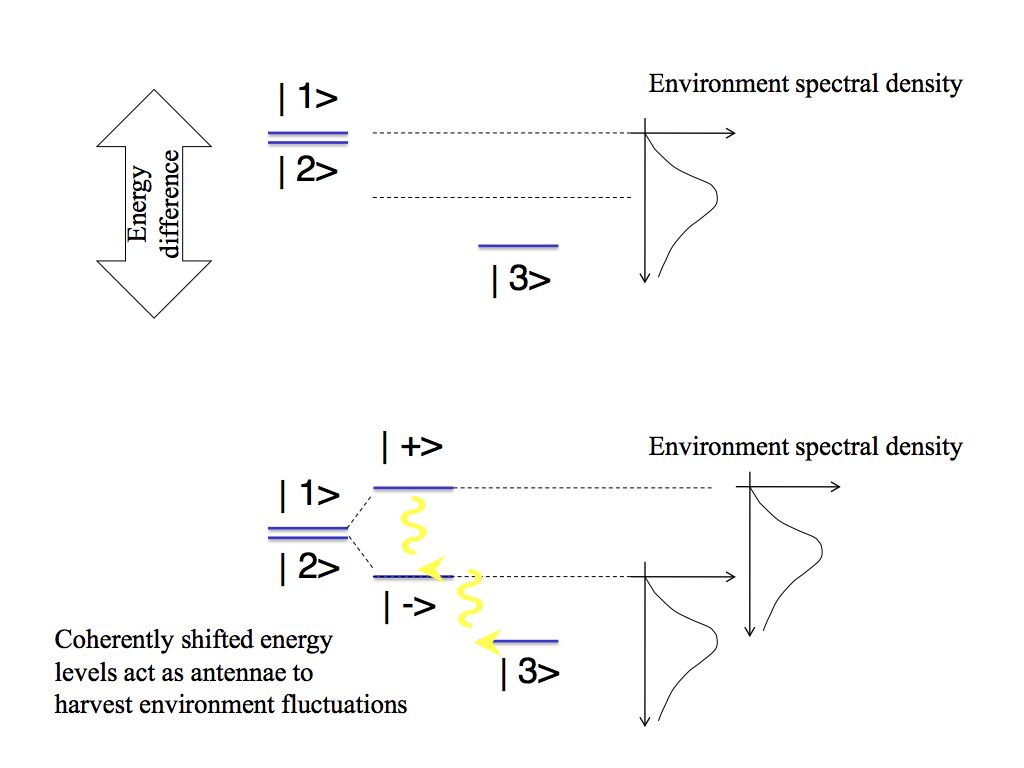}}
\caption{In the upper half of the figure, two closely spaced energy levels
are separated from a third level to which excitations should be delivered.
A coherent interaction between the upper two energy levels leads to dressed states $|\pm\rangle$ with an energy
splitting which, if matched to the maximum of the environment spectral
density will optimize transport from the upper to the lower level. Hence
the split levels act as an antennae for environmental fluctuations.}
\label{FigAntennae}
\end{figure}
\begin{figure}[hbt]
\centerline{\includegraphics[width=7.5cm]{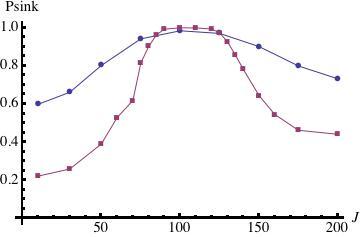}}
\caption{Population transferred to sink at $t=5 \mathrm{ps}$ as a function of $J_{12}$
for a Lorentzian spectral density with FWHM of $50 \mathrm{cm}^{-1}$ (blue dots) and
$10 \mathrm{cm}^{-1}$ (purple squares).  Numerical results obtained using secular
Bloch-Redfield equations. For all curves $J_{23}=30\mathrm{cm}^{-1}$ and the environment
temperature is $77$ K. }
\label{redfield}
\end{figure}
As an illustrative example consider again the toy model consisting in a three site network
where site 3 is coupled to a reaction center and let us assume that a Lorentzian spectral
density that is peaked at $200cm^{-1}$ (which is typical for FMO) and a FWHM of around
$50cm^{-1}$ (which is narrower than estimated for the FMO). Setting the energy of site
$1$ and $2$ to $\epsilon_1=\epsilon_2=300 cm^{-1}$ and $\epsilon_3=0cm^{-1}$, it then
turns out that the transfer will be optimized for a coupling $J_{12}$ between sites $1$
and $2$ of around $J_{12} = 100cm^{-1}$ which results in a splitting of $\Delta E = 200
cm^{-1}$. Figure \ref{redfield} shows the population transferred to a sink $P_{sink}$
connected to site $3$ after $5$ ps of evolution described by a standard secular Bloch-Redfield
master equation and the Lorentzian spectral function described above. The transfer to the
sink was described by a Lindbladian with a transfer rate of $1 \mathrm{ps}^{-1}$, $J_{13}=0$,
$J_{23}=30 \mathrm{cm}^{-1}$ and the bath temperature was taken as $77$K. A clear maximum is
seen for $J_{12}=100 \mathrm{cm}^{-1}$ which can be interpreted as a result of the phonon
antenna effect set out in Fig. \ref{FigAntennae}. Figure \ref{redfield} also shows results
for a narrower Lorentzian (FWHM $10 \mathrm{cm}^{-1}$), showing that the enhancement at
$J=100 \mathrm{cm}^{-1}$ is much more strongly peaked as one would expect.  These results
suggest that exciton couplings might be used to exploit the structure of the noise spectrum
of the protein surroundings, and that the advantages of doing this increase as the structure
of the environmental spectral function becomes sharper.

However, when the structure of the spectral function contains very sharp resonances, the
resonant modes can no longer be treated with standard master equation approaches. The derivation
of most master equations from microscopic system-bath Hamiltonians normally relies on the fact
that the environment contains an extremely large number of modes and that the coupling to any
individual mode is weak. However, a spectral function that is very sharply peaked at some
characteristic frequency $\omega$ implies a strong coupling to a discrete mode of the environment,
and such an interaction can induce a significant \emph{coherent} modification of the exciton
dynamics while the mode is driven strongly out of equilibrium by the exciton dynamics. These
discrete mode effects were explored using a recently-developed time-adaptive renormalisation 
group technique that allows strong interactions to discrete modes to be treated as part of 
an environment with an arbitrary residual spectral density \cite{PriorCH+2010}. This method
produces numerically exact results with controllable errors, and it was found that discrete modes
may \emph{coherently} modulate population dynamics over ps timescales whilst the continuous part
of the spectral density drives efficient population transfer \cite{PriorCH+2010,ChinRH+2010}.

Another approach to this problem is to consider the strong coupling to a discrete mode as
part of the system and then to treat the residual smooth parts of the spectral density as
the environment. This approach requires a large expansion of the Hilbert space of the effective
exciton-mode system and makes simulations considerably more complex, especially if the residual
bath has a non-Markovian character. This approach is thus only suitable for simple systems with
relatively simple residual spectral densities, but it is very useful for illustrating the
elementary concepts of strong exciton-mode interactions. We now consider such a model using the
Hamiltonian given by Eq. (\ref{ham}) to describe three sites, as described above, in which each
site is coupled coherently to a single harmonic oscillator of frequency $\omega=200 \mathrm{cm}^{-1}$
and coupling constant $g_{i}=30 \mathrm{cm}^{1}$. To clearly demonstrate the discrete phonon antenna
effect, we neglect background dephasing here, but this will be treated in detail in a forthcoming
work \cite{delReyCH+2012}. The simulations are performed using exact diagonalisation and the simulation
results were found to converge with $4$ bosonic levels per mode.

\begin{figure}[hbt]
\centerline{\includegraphics[width=7.5cm]{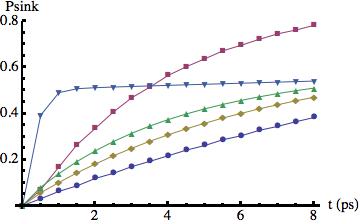}}
\caption{$P_{sink}$ as a function of time for three-site system, where each site is coupled to a mode of frequency $200 \mathrm{cm}^{-1}$ with coupling strength $g=30 \mathrm{cm}^{-1}$. A strong enhancement of transport is seen for $J_{12}=100 \mathrm{cm}^{-1}$ (squares). The other curves are $J_{12}=150 \mathrm{cm}^{-1}$ (diamonds),$J_{12}=200 \mathrm{cm}^{-1}$ (triangles),$J_{12}=300 \mathrm{cm}^{-1}$ (inverted squares) and $J_{12}=50 \mathrm{cm}^{-1}$ (dots). For all curves $J_{23}=30\mathrm{cm}^{-1}$ and the initial state of the vibrational mode is a thermal state at $77$ K.  }
\label{mode}
\end{figure}
\begin{figure}[hbt]
\centerline{\includegraphics[width=7.5cm]{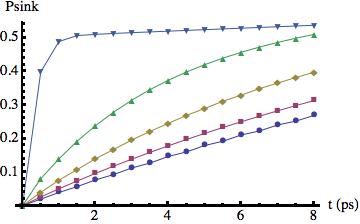}}
\caption{$P_{sink}$ as a function of time for a three-site system without any coupling to a mode. Weak transport is now seen for $J_{12}=100 \mathrm{cm}^{-1}$ (squares). the other curves are $J_{12}=150 \mathrm{cm}^{-1}$ (diamonds),$J_{12}=200 \mathrm{cm}^{-1}$ (triangles),$J_{12}=300 \mathrm{cm}^{-1}$ (inverted squares) and $J_{12}=50 \mathrm{cm}^{-1}$ (dots). For all curves $J_{23}=30\mathrm{cm}^{-1}$ and the initial state of the vibrational mode is a thermal state at $77$ K.}
\label{nonoise}
\end{figure}
\begin{figure}[hbt]
\centerline{\includegraphics[width=7.5cm]{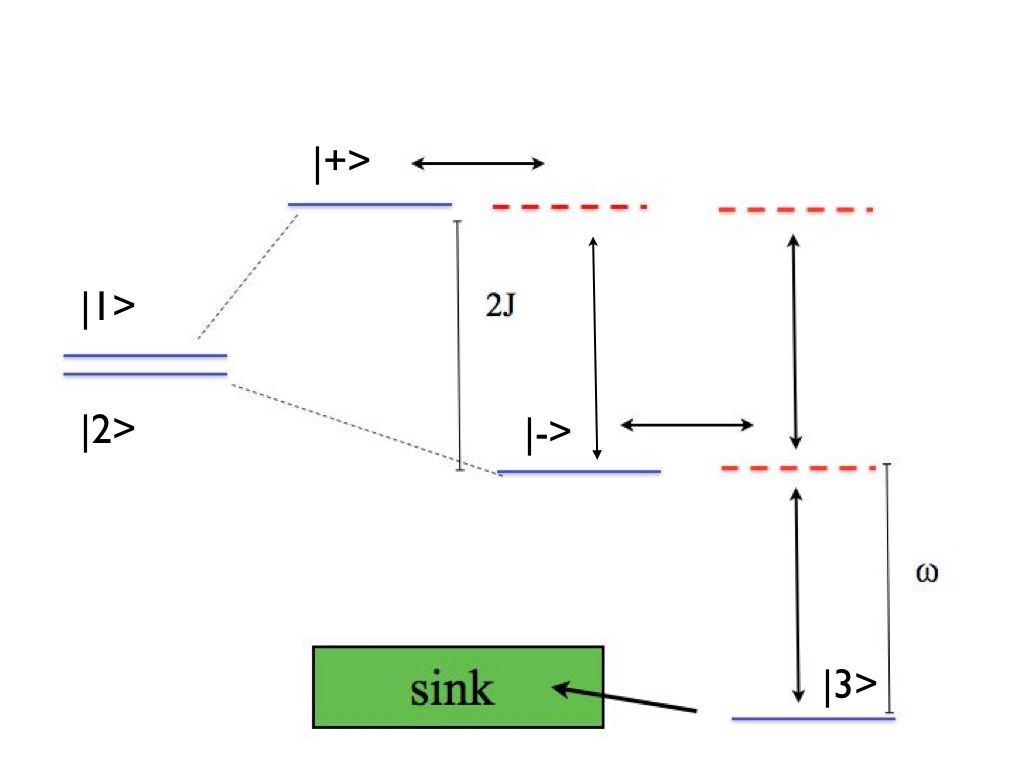}}
\caption{Energy level scheme for three-site and sink system (blue lines) showing vibrational sideband levels (dashed red lines). The coupling $J_{12}$ splits the degenerate transitions, forming new super position states $|\pm\rangle$,   and can bring these electronic levels into resonance with the sidband structure. Under these conditions fast, coherent inter-site transport (black arrows) becomes possible through the new exciton-mode energy level landscape.}
\label{antennamode}
\end{figure}

Figure \ref{mode} shows the population in the sink as a function of time for a range of $J_{12}$ values.
It can be seen that the behaviour of $P_{sink}$ is strongly non-monotonic with a large enhancement of the
population transport when $J_{12}=100 \mathrm{cm}^{-1}$. For these values the hybrid states $|\pm\rangle$
are resonant with the vibration sidebands of each site (see Fig. \ref{antennamode}) and the mode-exciton
interaction mediates strong transitions between sites by allowing efficient coherent tunneling between
the degenerate vibrationally excited states of the electronic excitations. For other values of $J_{12}$
the $|\pm\rangle$ states are detuned from the mode sidebands and vibration-mediated transport is dramatically
reduced. Note also that for $J_{12}=300 \mathrm{cm}^{-1}$ the lower state $|-\rangle$ is resonant with
site $3$. This leads to a rapid transport of $50\%$ of the population, but the $50\%$ initially held
in the $|+\rangle$ state is at such a high energy that it cannot transfer population to the sink via
site $3$ in the absence of background noise.

Figure \ref{nonoise} shows results for the same parameters in the absence of the mode. One now observes
a monotonic behaviour of the transport rates as $J_{12}$ is increased, and by comparison with Fig.
\ref{mode}, the dramatic resonant enhancement of transport observed at  $J_{12}=100 \mathrm{cm}^{-1}$
is highlighted. These results suggest that tuning of excitonic couplings to "harvest" the available noise
spectrum is most effective when the quasi-discrete features are exploited. It is in this sense - just as
excitonic interactions allow chromophores to sample larger parts of the optical spectrum - that we refer
to such coherent sampling of discrete features or maxima of the spectral function as the phonon antennae
effect.  Not only do these allow for tremendous gains in transport efficiency, their strong quantum
mechanical interaction with the exciton appears to strongly modify the behaviour and lifetime of coherence
dynamics (not shown).

\subsection{The principles at work: FMO and LHII}\label{fmo}
Here we will briefly analyse how these principles are put to work
to explain aspects of noise assisted transport in two different
pigment-protein complexes.

\subsubsection{The FMO complex}
To illustrate how the general principles outlined in the previous section
may feature in a realistic scenario, we will consider the example of
excitation transfer across a monomer of the FMO complex, which can be modeled
as a $7$ sites network and for which detailed information concerning the
system Hamiltonian is available \cite{AdolphsR2006}. We ignore here the
recent addition of an eigth site \cite{SchmidtanBuschME+2011} as it does not affect
the general arguments and was unlikely to have been present in recent
experiments reported in \cite{EngelCR+2007,PanitchayangkoonHF+2010}.

\begin{figure}[hbt]
\centerline{\includegraphics[width=7cm]{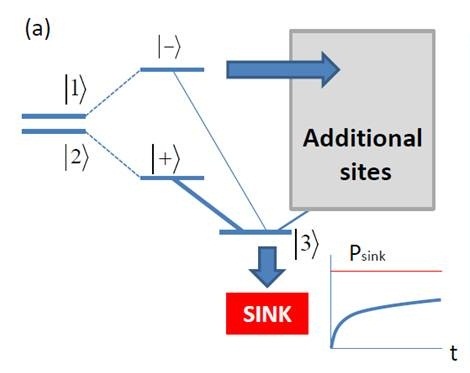}}
\centerline{\includegraphics[width=7cm]{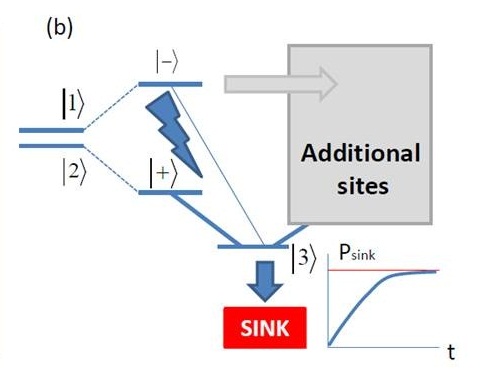}}
\caption{Noise-assisted energy transfer across FMO can be qualitatively understood
by introducing a hybrid basis of local sites. While a purely unitary evolution yields
to inefficient transport, as discussed in the text, the presence of dephasing noise
eliminates inefficient transport paths while opening up new channels for excitation
transfer.  While the estimation for the transfer efficiency under coherent evolution
is well below 100$\%$, a simple noise model brings this number close to perfect
efficiency within the observed transfer time.}
\label{Fig2}
\end{figure}

Given the strong coupling of sites $1$ and $2$, these levels are shifted
and mixed and the dynamics is conveniently described using an hybrid
basis \cite{ChinDC+2010} for these two sites that we denote by $|\pm\rangle$.
In this basis, the Hamiltonian has the local site energies and coupling
structure shown in Figure \ref{Fig2}a, where site $3$ is connected to a
sink node from which excitation is transferred to the reaction center
and all the remaining sites have been packed in a block (labeled as
"additional sites" in Figure 2) that is uncoupled from the level $|+\rangle$.

When an excitation is injected in site 1, the coherent evolution leads to
transfer efficiency $P_{sink}$ below $60\%$, far from the ideal transfer,
represented in Fig. \ref{Fig2} by the red horizontal line. The {\em coherent
interaction} between sites $1$ and $2$ leads to level splitting and moves
one of these energetically closer to site $3$ while the other is farther
removed. The introduction of dephasing noise with a given spectral density
can dramatically add to this picture and lead to perfectly efficient transport
to the sink, as illustrated in the part (b) of Fig. \ref{Fig2}. Environmental
fluctuations lead to transitions between states $|+\rangle$ and $|-\rangle$
and the resulting transition rate will depend on the spectral density at
the level splitting of the dressed states $|+\rangle$ and $|-\rangle$.
Indeed, inspection of the relevant Hamiltonian reveals that the
coupling between sites $1$ and $2$ is of the order of $93cm^{-1}$ and
leads to a splitting between the dressed states $|+\rangle$ and $|-\rangle$
of very close to $200cm^{-1}$ which matches the maximum of the spectral
density of the environmental fluctuations used in \cite{AdolphsR2006}.
It is hence tempting to speculate that nature has optimized the intersite
coupling and/or spectral density to provide a perfect match between the
maximum of the spectral density and the level splitting of the dressed
states $|+\rangle$ and $|-\rangle$ thus exploiting the idea that quantum
coherent interaction support an "antenna for phonons".

Furthermore, coherent oscillations between level $|-\rangle$  and the rest of
the complex are now largely suppressed (due to noise and destructive interference),
while an incoherent transfer path between previously decoupled levels $|\pm\rangle$
is now active, leading to a fast transfer of population to the sink via $|-\rangle$.
Within this model, it is possible to optimize the local dephasing rates so as to
reproduce the observed transfer rates in the correct time scale. More detailed
discussions can be found in the literature \cite{MohseniRL+2008,PlenioH2008,CarusoCD+2009,ChinDC+2010}.

Recent molecular dynamics simulations have also shown that the spectral density of
pigment-protein complexes such as the FMO complex contain a large number of sharp
features which could be exploited via the mechanism described above \cite{OlbrichSSK,OlbrichK2010,ShimRVAG}.
Moreover, spectral hole burning and fluorescence line narrowing experiments have also
shown that a number of modes with frequencies comparable to the energy differences
between excitons couple strongly to the electronic degrees of freedom in the FMO
complex \cite{wending,johnson,CaycedoSolerCA+12}.  Given that vibration-mediated tunneling is also
implicated in a range of other important biological transport processes, such as
olfaction \cite{brooks}, it does not seem unreasonable that EET dynamics in
pigment-protein complexes could take advantage of the rich spectral structure of
their local environments by appropriately tuning the inter-site interactions to
sample the maxima of the spectral function.

Indeed, it has also been shown in \cite{ChinDC+2010} that a discrete mode coupled to each
site in the FMO complex could also induce efficient transport, and in some cases
could even outperform the optimised, purely incoherent noise assisted transport which
arises from a Lindblad treatment of the environment. Given the
observed presence of a wide range of discrete features in the phonon spectrum of FMO,
it is likely that both incoherent and discrete phonon antenna effects could play a role
in the efficient energy transport seen in experiments, with the latter mechanism possibly
contributing to  enhanced, longer-lasting coherent dynamics, as observed in
\cite{ChinDC+2010}. A full analysis of the dynamics of FMO in the presence of both
discrete modes and continuous spectral densities with maxima close to exciton energies
differences will be appear in \cite{ChinHP+2011}.

\subsubsection{The LHII complex}
Here we consider the LHII complex from the photosynthetic unit of purple bacteria
\cite{HuRD+2002}. The photosynthetic unit is composed of a membrane that contains
only two types of antennae complexes, LHI and LHII. The former are believed to form
ring-like structures that enclose the reaction center while the latter transfer energy
to the reaction center via the LHI antennae. The LHII antennae consists of two subunits,
the B800 and the B850 ring. The B800 ring, absorbing around $800$nm, consists of nine
pigments, which have their molecular plane arranged perpendicular to the symmetry axis.
The B850 ring, absorbing around $850$nm, consists of nine repeating pairs of
$\alpha$-$\beta$ pigments, with opposite transition dipole moments and their molecular
planes parallel to the symmetry axis.

Here we make the simplifying assumption that the interaction between pigments is well
approximated by a dipole-dipole interaction of the form
\begin{equation}
    V_{ij} = \frac{1}{4\pi\epsilon |{\vec r}_{ij}|^3} \left[
    {\vec \mu}_i{\vec \mu}_j - 3({\vec \mu}_i{\hat {\vec r}_{ij}})
    ({\vec \mu}_i{\hat {\vec r}_{ij}})  \right]
\end{equation}
where ${\vec r}_{ij} = {\vec r}_i - {\vec r}_j$ is the vector joining
pigment $i$ with pigment $j$ and ${\hat {\vec r}_{ij}}$ is the corresponding
unit vector. We chose $\epsilon = 1.3\cdot 8.85 10^{-12} C^2/Nm^2$.
The dipole moments are assumed to be of equal modulus $|{\vec \mu}|
= 2.046\times 10^{(-29)} Cm$. The dipole moments are nearly perpendicular
to the symmetry axis and to the radial direction (see \cite{CogdellGK2006}
for details).

With this information it is possible to determine the interaction strengths
between any pair of pigments. In the B800 ring neighboring pigments experience
an interaction strength of around $-26.7cm^{-1}$ while in the B850 ring
neighboring pigments within a $\alpha$-$\beta$ pair experience an interaction
of about $190.8cm^{-1}$ and between neighboring pairs of $239.4cm^{-1}$
while the coupling from B800 to B850 between the closest pigments is
of the order of $12cm^{-1}$. Employing these parameters it is now straightforward
to observe that the transfer from the B800 to the B850 ring benefits from
the presence of dephasing noise.
\begin{figure}[hbt]
\centerline{\includegraphics[width=9.5cm]{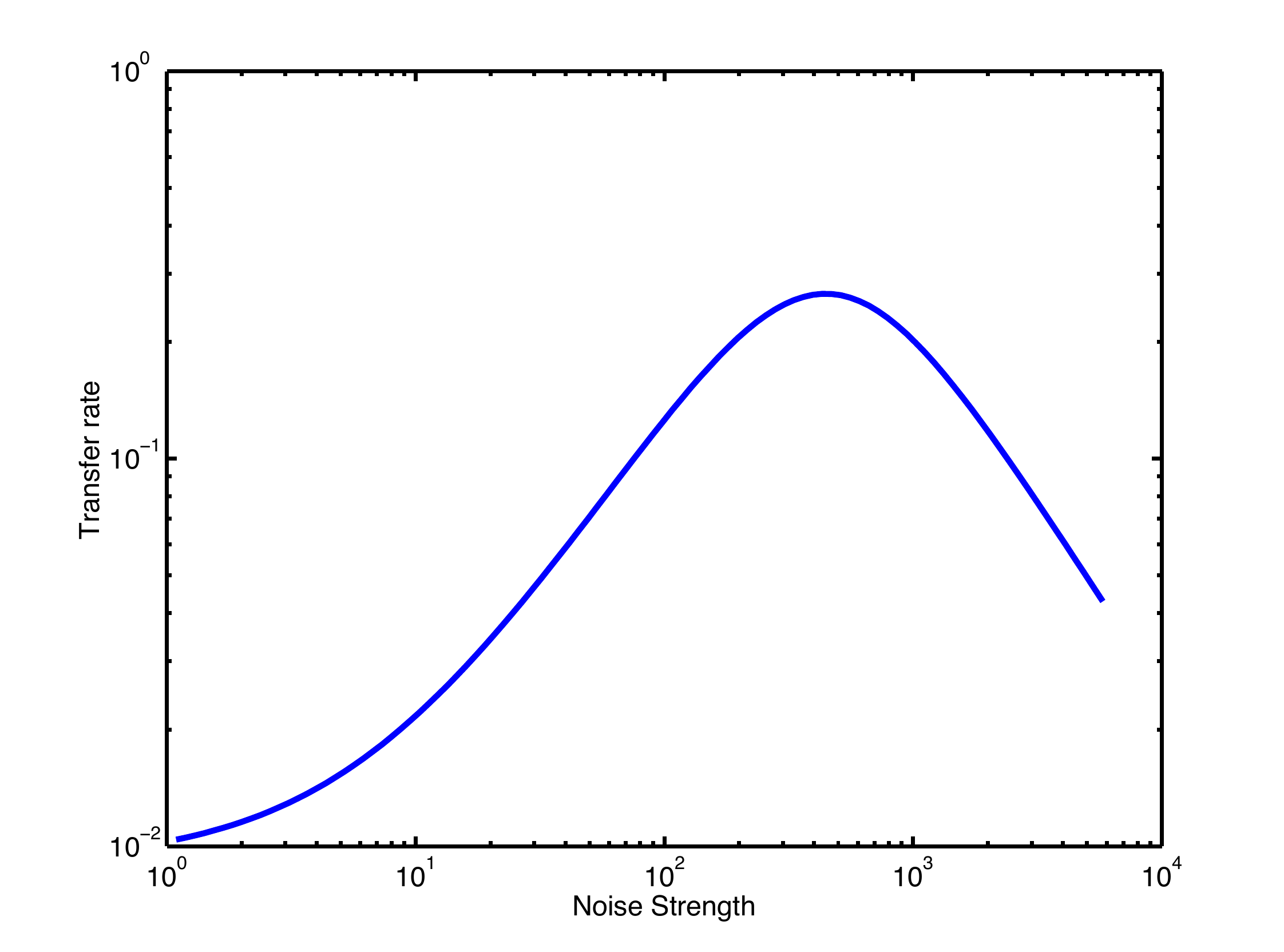}}
\caption{The B800-B850 transfer rate versus the noise strength exhibiting
a non-monotonic behaviour in the noise strength.}
\label{fig4}
\end{figure}
Fig \ref{fig4} shows the non-monotonic
dependence of the transfer rate between the two rings versus the noise
strength under the assumption that all pigments are subject to noise of identical
strength. Here the noise assists in bridging the energy gap between the
B800 and the B850 ring. Strong noise does however tend to suppress delocalization
of excitons in each ring and thus suppress coherent superpositions that
may lead to enhanced transport from B800 to B850.

\section{Reflections on the microscopic origin of long-lived coherence}
Having given an overview of the key concepts underpinning noise-assisted transport, we
proceed with our general discussion of the role of noise in PPC dynamics by going a little
more into the details of the microscopic couplings which generate the essential processes
of pure dephasing (destruction of interferences) and exciton relaxation. Without leaping
into major simulations, we aim to point out a number of effects which might be observable
in experiments, and which could potentially help us to pinpoint the physics behind a number
of currently ill-understood observations that have been made in  experiments and ab initio
simulations.

Let us consider the following standard Hamiltonian of chromophores which are linearly coupled
to protein fluctuations in their excited states. Defining bosonic annihilation and creation
operators $a_{ki},a_{ki}^{\dagger}$  for the independent, uncorrelated fluctuations of frequency
$\omega_{k}$ of the protein environments acting on site ${i}$, and denoting an optical excitation
on site $i$ by the state $|i\rangle$, the Hamiltonian for a single excitation can be written as,

\begin{eqnarray}
H&=& \sum_{i}\left((\epsilon_{i}+X_{i})|i\rangle\langle i| +H^{B}_{i}\right)+\sum_{i\neq j}J_{ij} |i\rangle\langle j|,\label{ham}
\end{eqnarray}
where $X_{i}=\sum_{k}g_{ik}(a_{ki}+a_{ki})$ and $H^{B}_{i}=\sum_{k}\omega_{k}a_{ki}^{\dagger}a_{ki}$, where $\omega_{k}$ are the frequencies of the protein fluctuations and $g_{ik}$ their coupling strength to state $|i\rangle$ ($\hbar=1$ throughout). Assuming identical environments for each site, the dissipative effect of the environment on the excitons is fully-characterised by the spectral density of the environment $G(\omega)=\sum_{k}g_{ik}^{2}\delta(\omega-\omega_{k})$, which we discussed extensively in the previous section. It is normally taken that the electronic couplings $J_{ij}$ between sites is stronger than the pigment-protein interaction and that the appropriate basis for analysing experimental results corresponds to that of the excitonic eigenstates $|e_{n}\rangle=\sum_{i}C^{n}_{i}|i\rangle$.

Rewriting the Hamiltonian in the exciton basis, we obtain,
\begin{eqnarray}
H&=&\sum_{n}E_{n}|e_{n}\rangle\langle e_{n}|+ \sum_{n}Q_{n}|e_{n}\rangle\langle e_{n}|\nonumber\\
&+& \sum_{n\neq m}(R_{nm}|e_{n}\rangle\langle e_{m}|+R_{nm}^{\dagger}|e_{m}\rangle\langle e_{n}|\nonumber\\
&+&\sum_{i}H^{B}_{i}.\label{exham}
\end{eqnarray}
The new couplings to the bath are given by $Q_{n}=\sum_{k,i}g_{ik}|C^{n}_{i}|^{2}(a_{ki}+a_{ki}^{\dagger})$ and $R_{nm}=\sum_{k,i}g_{ik}C^{n}_{i}C^{m}_{i}(a_{ki}+a_{ki}^{\dagger})$. As was
mentioned in the introduction, in this basis the exciton states,  which is
the appropriate basis for weak system environment coupling, are stationary
under the action of the electronic Hamiltonian and dynamics arise solely
through the action of the bath couplings on the system. From Eq. (\ref{exham})
one can see that these interactions come in two distinct flavours, namely
inter-exciton (transverse) transfer terms ($\propto R_{nm}$) and diagonal
(longitudinal) fluctuations of the exciton energies ($\propto Q_{n}$). During
the evolution of an initially-prepared superposition of exciton states, the
total dephasing time of the superposition roughly contains two contributions,
a relaxation part coming from transverse processes and a pure dephasing part
coming from the longitudinal processes.

Before going on to discuss the case
of uncorrelated fluctuations, we briefly present the parameters
\begin{eqnarray}
    \gamma_{nm}&=&\sum_{i,j}C^{n}_{i}C^{m}_{i}C^{n}_{j}C^{m}_{j}e^{-r_{ij}/r_{c}},\\
a_{nm}&=&\sum_{i,j}((C^{m}_{i})^2(C^{m}_{j})^2+(C^{n}_{i})^2(C^{n}_{j})^2\nonumber\\
&-&2(C^{m}_{i})^2(C^{n}_{j})^2)e^{-r_{ij}/r_{c}},
\end{eqnarray}
where $r_{c}$ is the correlation length of the fluctuations at each site and $r_{ij}$ are
the inter-pigment distances  \cite{AdolphsR2006}. The off-diagonal quantities $\gamma_{nm}$ are proportional to
the strength of the transition rates between excitons $n$ and $m$ in Bloch-Redfield theory, and pure-dephasing processes between excitons $n$ and $m$ in modified Redfield theory are proportional to $a_{nm}$  \cite{AdolphsR2006}. Although
we will not discuss spatial correlations explicitly, we note that when $r_{c}$ is much larger
than the size of the complex $\gamma_{nm}\approx |\langle e_{n}|e_{m}\rangle|^2$, $a_{nm} \approx 0$ and the
\emph{total} (transverse and longitudinal) coupling to the bath \emph{vanishes} in the
one-excitation manifold. Spatial correlations thus reduce dephasing \emph{and} transport
rates by roughly similar amounts.

Returning to uncorrelated environments, it is well known that delocalisation of
the excitons over extended regions of roughly $N$ sites leads to a $1/N$ suppression
of the pure-dephasing of ground state-excited state coherences via the phenomena of \emph{motional
narrowing} \cite{cho}. However, due to disorder the delocalisation length in typical
PPCs is often restricted to only a few sites, and indeed several PPCs \emph{only}
contain a few sites in total. Nevertheless, even if the excitons are only delocalised
over two sites, a significant suppression of zero-quantum coherence dephasing  can
arise if the electronic wavefunctions are nearly equally distributed over the two
sites, i.e. the eigenstates are symmetric and anti-symmetric combination of the
two local excitation states. To see this clearly, imagine we have a dimer system
in which $\epsilon_{1}-\epsilon_{2}\ll J$. The two excitonic states are then
$|e_{\pm}\rangle=(|1\rangle\pm|2\rangle)/\sqrt{2}$. In the exciton basis the
exciton-phonon Hamiltonian becomes,

\begin{eqnarray}
H&=&\sum_{n=+,-}E_{n}|e_{n}\rangle\langle e_{n}|+ \frac{1}{2}Q_{n}\left(\underbrace{|e_{+}\rangle\langle e_{+}|+|e_{-}\rangle\langle e_{-}|}_{\mathrm{=i.d}}\right)\nonumber\\
&+& (R_{+-}|e_{+}\rangle\langle e_{-}|+ \mathrm{h.c})+\sum_{n}H^{B}_{n}.\nonumber\\
\end{eqnarray}
As indicated by the underbrace, the longitudinal coupling to the excitons vanishes (in the one-excitation sector) when the eigenstates are symmetric/anti-symmmetric combinations. The bath-induced dynamics now proceeds purely through transverse terms and dephasing only arises through relaxation. We shall refer to this case as \emph{relaxation-limited dephasing}. In general, if the excitons are delocalised over $N$ sites and excitonic couplings are very strong, then the eigenstates are of the form $|\tilde{e}_{n}\rangle=\sum_{k=1}^{N}e^{-ink/N}|i\rangle$ and will also not be subject to any longitudinal bath coupling.

Our discussion in this section has so far made no assumptions about the spectral function of the environment, but in order to illustrate a few ideas we will now consider the case of a sufficiently weak exciton-phonon coupling where the influence of the bath can be treated in a Markovian (Redfield) approximation. In this limit, one can rigourously define the relaxation time $T_{1}$, the total dephasing time of coherences $T_{2}$ and the pure-dephasing time $T_{2}^{*}$. The three times are related by $1/T_{2}=1/2T_{1}+1/T_{2}^{*}$.  The relaxation time $T_{1}^{-1}$ describes the rate of population transfer induced by the transverse couplings in Eq. (\ref{exham}) and $T_{2}^{*}$ arises from longitudinal terms.  In the limit of relaxation-limited dephasing, where $T_{2}^{*}$ diverges,  $T_{2}=2T_{1}$ and the dephasing rate is actually \emph{slower} than the population transfer rate. However, in most experiments performed in liquids and the solid state, $T_{2}\ll T_{1}$ due to the dominant contribution of pure dephasing terms. This is often also true in realisations of qubits in quantum information science, and removal of these processes is currently a topic of major research.

In current 2D spectroscopy experiments it is often possible to meaure the population lifetimes (giving an estimate of $T_{1}$) and the total dephasing time ($T_{2}$) of coherences seperately. By comparing these numbers, the extent of pure dephasing can be inferred and the extent of the suppression of pure-dephasing terms by delocalisation might also be extracted. This might possibly allow strongly-coupled excitonic states to be identified and could also thereby provide a consistency-check of Hamiltonians and dynamical models which have been proposed for these systems. From this point of view, it must be noted that non-Markovian dynamics are thought to play an important role in PPCs such as FMO \cite{IshizakiF+2009}, and in these cases there is no simple relation between the typical timescales associated with pure dephasing, relaxation and the total dephasing timescale(s). However, non-perturbative simulation techniques can address this, and this will be dealt with in a forthcoming work \cite{ChinHP+2011}.

One example of this effect could be found by looking at excitons associated with sites $1$ and $2$ of FMO. In most published Hamiltonians the difference between local excitation energies of sites $1$ and $2$ are smaller than the large excitonic coupling between then, and two exciton states contain strong contributions from symmetric and anti-symmetric combinations of sites $1$ and $2$. Within the model and modified Bloch-Redfield approximation presented by Adophs and Renger \cite{AdolphsR2006}, the coherence $\rho_{nm}$ between excitons $n$ and $m$ decoheres under pure dephasing processes as $\rho_{nm}\propto e^{-a_{mn}(g(t)-g(0))}$ ( we have neglected the relaxation-induced contribution to the decoherence). The coherence between exciton $n$ and the ground state $\rho_{gn}$ similarly decoheres as $\rho_{gn}\propto e^{-\gamma_{nn}(g(t)-g(0))}$. In both cases, the spectral density-dependent functions $g(t)$ are the same \cite{AdolphsR2006}. The relative strengths of the pure dephasing dynamics of the inter-exciton and ground-exciton can be therefore be estimated by looking at the ratios $a_{nm}/\gamma_{nn}$ or $a_{nm}/\gamma_{mm}$, which are independent of the spectral function.   Lower and upper estimates of these ratios were obtained by computing $a_{nm}, \gamma_{nn}$ and $\gamma_{mm}$ for the excitons $|e_{n}\rangle, |e_{m}\rangle$ of  FMO Hamiltonians with the largest contributions from sites $1$ and $2$.  Assuming uncorrelated environments, we found that $ a_{nm}/\gamma_{nn}\approx a_{nm}/\gamma_{mm}\approx 10^{-1}-10^{-3}$, with the lower bound being obtained from the Hamiltonian of Adolphs and Renger \cite{AdolphsR2006} and the upper bound from the Hamiltonian presented in \cite{Hayes2011}.  The large suppression of the inter-exciton pure dephasing contribution at the upper bound, would most likely lead to a relaxation-limited dephasing.

After completion of this work, an explicit example of relaxation-limited dephasing has also been given in \cite{brumer}. They have computed the dynamics for site population dynamics of sites $1$ and $2$ of the FMO complex using a weak coupling non-interaction blip approximation (NIBA) developed in the context of the spin-boson problem \cite{weiss}. Pachon and Brumer \cite{brumer} point out for the parameters of these sites, the small site energy difference and large excitonic coupling leads to a suppression of pure-dephasing contributions to the total dephasing rate, and at $77$ K, the dephasing rate they predict is indeed close to the relaxation-limited case $T_{2}\approx 2T_{1}$. However, as we stress above, this relaxation-limiting suppression of pure-dephasing - which does not rely on a non-Markovian treatment of the system-bath interactions, is only relevant under conditions where site energies differences are much smaller than inter-site couplings. In the FMO complex experiments, the long-lasting excitonic coherences are observed between the two lowest energy excitons, which correspond to sites $3$ and $4$, rather than sites $1$ and $2$. These sites have energy differences and couplings of roughly equal magnitude \cite{AdolphsR2006}, or much greater site energy differences \cite{Hayes2011}. While relaxation-limited dephasing can play a role for certain pathways in the  FMO dynamics, it is unlikely to provide a \emph{general} explanation for the most significant long-lasting coherences observed in FMO, as is claimed in \cite{brumer}. It should also be pointed out that the long-lasting inter-exciton (not to be confused with inter-site) coherences at $77$ K in FMO persist for approximately $1.8$ ps, much longer than those predicted in \cite{brumer}.

It is important to also note that the long-lasting coherences found in \cite{brumer} are not due to any non-markovian effects; the use of the NIBA approach (which may not be reliable in the relevant parameter regime, see footnote \cite{footnote1} gives non-perturbative \emph{renormalisation} of the parameters which determine the relaxation and dephasing rates, but the dynamics themselves are simple, strictly exponentially-damped oscillations and the coherence lifetimes could be predicted from polaron perturbation theory. This, and the agreement of this theory with numerical simulations, could also suggest that for the simple Ohmic spectral density and weak-coupling parameters used in the literature, advanced simulation techniques are not required for complexes such as FMO, though a comprehensive understanding of EET dynamics is still lacking as it is unlikely that the simple Ohmic spectral density describes all aspects of the system-bath physics.

One of our reasons for discussing delocalisation-induced suppression of pure-dephasing is that the interexciton coherences in FMO appear to dephase much more slowly  ($\sim 1$ ps) than the typical ($\sim70-150$ fs) dephasing times of the ground-exciton  coherences between excited and ground states \cite{PanitchayangkoonHF+2010}.  If the dephasing of ground-exciton coherences comes from the same set of fluctuations, one would expect roughly similar pure-dephasing times for the zero-quantum coherences as well. One potential solution is that the fluctuations are \emph{spatially correlated}, and has been used to analyse the results of a number of experiments on PPCs and conjugated polymers \cite{lee,collini}. However, a number of recent quantum chemistry and molecular dynamics simulations of the FMO protein have, somewhat inconveniently, found no evidence for significant spatial correlations between different pigments \cite{OlbrichSSK,OlbrichK2010,ShimRVAG}.

An alternative possibility that may be appropriate for excitons which are strongly delocalised over a few sites in PPCs is suggested by relaxation-limited dephasing. We have shown that strong excitonic coupling can suppress pure dephasing amongst excitons in the one-excitation sector, but when the electronic ground state $|g\rangle$ is also included in the Hamiltonian, it can be seen that the excitons are no longer protected from the longitudinal terms if they are in a state that spans excitation number sectors e.g. a ground-exciton coherence $|e_{n}\rangle\langle g|$.  This means that a ground-exciton coherence will feel a motionally-narrowed, but still potentially large pure dephasing, whereas an inter-exciton coherence between $|e_{\pm}\rangle$-like states will not be subject to this pure dephasing at all.

It is important to notice that, unlike spatial correlations which lead to a dynamical decoupling of the entire exciton-phonon interaction,  the relaxation-limited case corresponds to the entire weight of the exciton-phonon coupling being transferred to the inter-exciton couplings. In this sense, and under appropriate conditions, coherences are protected without effecting transport efficiency. The inter-exciton coherence dephasing time is now essentially determined by the lifetimes of excitons, and has a much stronger, and thus tunable, dependence on the exciton energy differences compared to pure dephasing which is frequency-independent. For example, the Redfield relaxation rate $\Gamma_{+-}$ between excitons $+$ and $-$ is given by \cite{AdolphsR2006},
\begin{equation}
\Gamma_{+-}=\pi[(1+n(E_{+-})G(E_{+-})+n(E_{-+})G(E_{-+})],
\label{gamma}
\end{equation}
where $E_{nm}=E_{n}-E_{m}$, $E_{n}$ is the energy of the $n$th exciton and $n(\omega)$ is the thermal bosonic occupation of a mode with frequency $\omega$. The total dephasing time of an inter-exciton coherence can be significantly longer than the ground-exciton dephasing time in the relaxation-limited case when the exciton lifetimes determined from Eq. (\ref{gamma}) are smaller than the ground-exciton dephasing rate. For our dimer example, when $\delta E_{\pm}\approx 2J$ is much larger than the maximum of $G(\omega)$,  the lifetimes of the states will be long because $\Gamma_{nm}\propto G(E_{nm})$.  The essential point here is that relaxation and pure-dephasing are determined by very different regions \emph{and} properties of the spectral function, with pure dephasing rates at long times being very sensitive to the \emph{functional form} of $\lim_{\omega\rightarrow 0} G(\omega)\coth(\beta\omega/2)/\omega^2$ \cite{ChinHP2+2011}. Therefore for transitions between states separated by large energy differences, relaxation-limited dephasing may significantly increase the effective inter-exciton dephasing time compared to the ground-exciton dephasing rates, this the difference depending sensitively on the spectral density used.

The mechanism described above is another example of how excitonic interactions can be used to shift the dominant effects of the environment between spectral features - here removing effects of aggressive low frequency dephasing fluctuations - and conceptually allowing lifetimes and coherence times on the same order of magnitude as each other and the transport time (which is controlled by the lifetimes). This could be an important \emph{part}, though certainly not the only mechanism, of the physics underlying long-lasting coherernces. However, the question of \emph{why} these systems might choose to do these things - especially the preservation of coherences - remains as elusive as ever.

We finally note that this idea of electronic protection of inter-exciton coherences could be tested in dimer systems with strong excitonic interactions such as those found in reaction centers, and a detailed theory of the ideas presented in the last two section of this article will appear shortly \cite{ChinHP+2011}.

\section{Conclusions}
In this work we have summarized recent work concerning several aspects of the dynamics
of quantum networks in contact with environments such as pigment-protein complexes. We
have discussed the basic principles underlying the dynamics of such systems and formulated
the concept of efficient transport mediated by phonon antennaes induced by coherent dynamics. We
also discussed the possible relevance of motional narrowing on the unexpectedly long
lifetime of zero coherences in pigment-protein complexes such as FMO.
A more detailed account of these two aspects will be published elsewhere.

\acknowledgements
This work was supported by the Alexander von Humboldt-Foundation, the EU
STREP project CORNER and the EU Integrated Project Q-ESSENCE. Aspects of
this work have benefitted from discussions with J. Cai, F. Caruso,
F. Caycedo-Soler, J. Prior, A. Datta, and G.S. Engel.


\begin{thebibliography}{99}
\bibitem{footnote1} We also note that the NIBA approach used in \cite{brumer} has been applied in a
rather different parameter regime than that in which the NIBA is derived, as for Ohmic environments
the usual condition for NIBA is that the parameters are in the scaling limit in which the cut-off
frequency of the bath $\omega_{c}$ is much larger than all other energy scales in the system. In
\cite{brumer}, the tunneling splitting is actually \emph{larger} than the bath cut-off, which makes
the subsequent use of the weak-coupling, biased Markovian theory unreliable, as there is no real
separation of system and bath time scales in the problem. Furthermore, the coupling strength used
is also not very far into the weak coupling limit required for this theory to be applied.
\end{thebibliography}
\end{document}